# Wave Packet Propagation through Graphene with Square and Triangular Patterned Circular Potential Scatterers

G.M. Milibaeva[1], H.T. Yusupov[2], D.G. Berdiyorova[3], Y. Rakhimova[4,5], M. Yusupov[6,7], A. Chaves[8], and Kh. Rakhimov[1,9,*]

[1] Institute of Materials Science, Uzbekistan Academy of Sciences, Tashkent 100084, Uzbekistan
[2] Kimyo International University in Tashkent, Tashkent 100121, Uzbekistan
[3] Liverpool John Moores University, Oryx Universal College, Doha, Qatar
[4] National University of Uzbekistan, Tashkent 100174, Uzbekistan
[5] Tashkent International University of Education, Tashkent 100207, Uzbekistan
[6] Institute of Fundamental and Applied Research, National Research University TIIAME, Tashkent, Uzbekistan
[7] Tashkent State Technical University, Tashkent 100095, Uzbekistan
[8] Universidade Federal do Ceará, 60455-900 Fortaleza, Ceará, Brazil
[9] Central Asian University, Tashkent 111221, Uzbekistan
* e-mail: kh.rakhimov@gmail.com

## Abstract

In this study, using the Dirac continuum model combined with the split-operator technique, we investigate the propagation dynamics of wave packets in graphene in the presence of circular potential barriers arranged in square and triangular geometries. Our results reveal a non-monotonic dependence of the wave packet transmission on the number of barrier rows along the propagation direction: the transmission initially decreases as rows of barriers are removed, but then increases again when additional rows are eliminated. To explain the observed nonlinear behavior, the time evolution of the transmission probability is analyzed, providing insight into the interplay between wave packet dynamics and the spatial arrangement of potential barriers.

These findings offer a pathway for designing graphene-based devices with tunable transport properties through engineered potential landscapes.

## INTRODUCTION

Graphene has emerged as a groundbreaking material due to its exceptional mechanical strength, high thermal conductivity, and outstanding electronic properties, including ultrahigh carrier mobility [1–5]. Its unique two-dimensional honeycomb lattice results in low-energy quasiparticles that behave as massless Dirac fermions, giving rise to unusual quantum phenomena such as Klein tunneling, wherein charge carriers can traverse potential barriers without backscattering [4]. While this effect is fundamentally intriguing, it poses significant challenges for nanoelectronic applications, where confinement and control of charge carriers are essential [2]. Therefore, a deeper understanding of how graphene responds to engineered potential landscapes is crucial for advancing its integration into practical nanoelectronic devices.

Extensive research has been devoted to understanding charge carrier dynamics in graphene, particularly in disordered environments featuring randomly placed scattering centers [3–6]. Such disorder has been shown to strongly influence transport by reducing transmission efficiency and increasing dwell times due to intensified scattering. In contrast, studies involving periodic arrangements of potential barriers have revealed a wide range of behaviors, including non-monotonic shifts in transmission as a function of barrier height and a pronounced dependence on the spatial configuration of the barriers [7–9]. Asymmetric or staggered periodic patterns, in particular, have demonstrated stronger scattering effects and greater suppression of transmission compared to symmetric lattices [12], emphasizing the importance of structural symmetry in modulating electronic transport. Additional investigations have explored wave packet propagation in ideal graphene, doped configurations [13,14], and under uniform magnetic fields [15–19]. Non-relativistic aspects of wave packet dynamics have also been explored in various material systems [8–12]. Despite these advances, the theoretical exploration of electron scattering by patterned potential barriers in graphene continues to be a vibrant and evolving area of study [20–26].

In this study, we investigate the propagation dynamics of wave packets in graphene under the influence of circular potential scatterers arranged in square and triangular lattices. Such configurations provide a promising platform for manipulating charge carrier behavior and enabling novel functionalities in graphene-based nanoelectronic devices. Utilizing the Dirac continuum model combined with numerical simulations, we analyze how transmission characteristics are influenced by key factors, including the height and radius of the potential barriers, as well as the presence of linear defects. Specifically, we examine transmission probabilities in systems with one- and two-line defects embedded within square and triangular lattices of circular barriers. By drawing analogies to classical wave propagation in two-dimensional periodic media, we extend this concept to the relativistic regime of graphene, offering a new perspective on charge carrier control through structural design. This approach resonates with prior research demonstrating the impact of engineered geometries on transport phenomena [29,30]. In addition to enriching our understanding of wave packet scattering in graphene, the proposed framework offers practical insights into tailoring electronic properties through spatial patterning. Building upon our previous work [6,7,21,31–33] and maintaining a consistent computational methodology, we further explore the capability of periodic potential landscapes to regulate transport behavior in Dirac materials.

## MODEL SYSTEM AND THEORETICAL FRAMEWORK

Our model system consists of a monolayer graphene sheet with lateral dimensions of $L = 1024$ nm (length) and $W = 128$ nm (width), into which circular electrostatic potential scatterers are embedded in regular square and triangular lattice arrangements. A schematic representation of the system is provided in Figure 1. Two configurations are considered: (a) **Sample 1** features a square lattice of circular barriers, representing a basic periodic potential landscape for benchmarking the influence on wave packet propagation; (b) **Sample 2** implements a triangular lattice of similar scatterers, offering a denser and geometrically distinct configuration for comparison.

The center-to-center spacing between adjacent potential barriers is set to $d = 16$ nm, and we consider two representative barrier radii, $R = 3$ nm and $R = 7$ nm, to investigate the influence of obstacle size on wave packet dynamics. To examine the effect of structural perturbations, we introduce linear defects by systematically removing one or two rows of barriers along the wave packet propagation direction. These defect lines, indicated by shaded regions in Figure 1, mimic either accidental imperfections or deliberate design features in a patterned potential landscape.

Such variations are expected to enhance scattering interactions and modulate the interference patterns generated during wave packet evolution. This setup enables a comprehensive investigation of how two-dimensional periodic arrays of potential barriers—with and without linear defects—affect charge carrier transmission, offering valuable insights into tunable transport in graphene-based nanostructures.

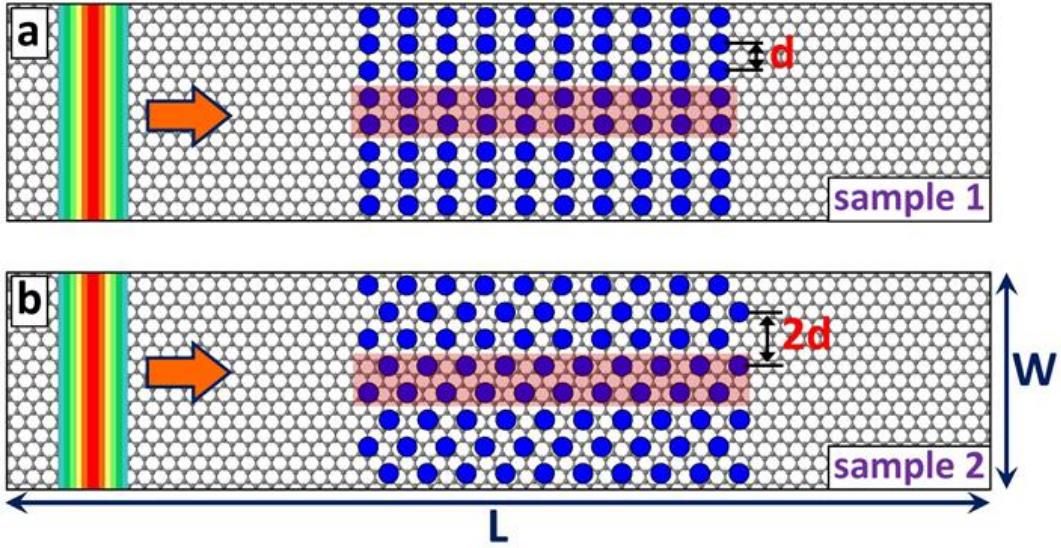

**Figure 1.** Schematic representation of wave packet propagation (orange arrow) in a graphene sheet with periodic circular potential barriers of positive value, arranged in (a) square and (b) triangular lattices. The wave packet is injected from the left. The shaded region highlights a line of potential barriers that can be selectively removed to investigate its effect on the wave packet propagation. Parameters L, W, and d indicate the system length, width, and barrier spacing, respectively. These configurations are labeled as **Sample 1** and **Sample 2**.

The wave packet is modeled as a Gaussian wave front characterized by its energy $E$ and width $a_x$. Specifically, the initial wave packet is a Gaussian function constant in the *y*-direction, with a finite width in the *x*-direction, and is expressed as:

$$\Psi(x,y,0) = N \begin{pmatrix} 1 \\ 1 \end{pmatrix} e^{ikx - \frac{x^2}{2a_x^2}} \qquad (1)$$

where $N$ is the normalization constant and $k=E/\hbar v_F$, with $v_F$ being the Fermi velocity in graphene. The pseudo-spinor $(1\ 1)^T$ is selected to ensure propagation along the *x*-direction, with $\langle\sigma_x\rangle=1$ and $\langle\sigma_y\rangle=\langle\sigma_z\rangle=0$. All simulations in this study assume a wave packet energy of $E=100$ meV and a width of $a_x=10$ nm. Wave packet propagation is governed by the time-evolution operator applied to the initial wave packet:

$$\Psi(x,y,t+\Delta t) = e^{-\frac{i}{\hbar}H\Delta t}\Psi(x,y,t) \qquad (2)$$

where $H$ is the Hamiltonian for low-energy electrons in graphene:

$$H = v_F\left(\vec{\sigma}\cdot\vec{p}\right) + V(x,y)\,\mathbf{I}, \qquad (3)$$

with $\vec{\sigma}$ representing the Pauli matrices, $\mathbf{I}$ being the 2×2 identity matrix, and $V(x, y)$ denoting the scalar potential. The wave functions are written as pseudo-spinors $\Psi=(\Psi_A,\Psi_B)^T$, where $\Psi_A$ and

($\Psi_B$) correspond to probabilities of the electron being in sublattices A and B, respectively. To simplify calculations, the split-operator technique [34,35] is employed:

$$\exp\left[-\frac{i}{\hbar}H\Delta t\right] = \exp\left[-\frac{i}{2\hbar}V(x,y)\mathbf{I}\Delta t\right] \exp\left[-\frac{i}{\hbar}v_F\vec{p}\cdot\vec{\sigma}\Delta t\right] \exp\left[-\frac{i}{2\hbar}V(x,y)\mathbf{I}\Delta t\right], \quad (4)$$

where terms of order $O(\Delta t^3)$ and higher are neglected. This approach enables efficient multiplication in real and reciprocal spaces, avoiding the explicit differentiation of the momentum operator by utilizing the Fourier transform and expressing $\vec{p}=\hbar\vec{k}$. Furthermore, the exponentials involving Pauli matrices can be exactly reformulated as matrices [31]. Simulations are performed with a time step of $\Delta t = 0.1$ fs, and the probabilities of finding the electron before and after the scattering region are computed. The latter is interpreted as the transmission probability ($P$) through the scattering region. To avoid numerical artifacts associated with periodic boundary conditions, we employ a sufficiently large computational grid of 128 nm × 1024 nm, while the scattering centers are confined to a smaller region of 128 nm × 128 nm around the x = 0 axis. This ensures accurate evaluation of the transmission and reflection probabilities, by allowing one to collect data on transmission probabilities before the wave packet reaches the edges of the computational box. All simulations are conducted at a wave packet energy of $E=100$ meV to ensure that the results remain within the range of applicability of the Dirac continuum model, i.e., around Dirac cone (low energy state).

## RESULTS AND DISCUSSION

As a key result, Figure 2 presents the transmission probabilities of the gaussian-shaped wavefront (see Eq. (1)) propagating in graphene, interacting with square (panels a, b) and triangular (panels c, d) lattices of circular potential barriers. The results are shown for two barrier radii: $R = 3$ nm (a, c) and $R = 7$ nm (b, d), plotted as a function of the barrier height $V_0$. We first examine the transmission probabilities for the defect-free configurations, indicated by the black curves in all panels of Figure 2. These represent the baseline transport behavior through perfectly ordered square and triangular lattices of circular potential barriers. For both radii ($R = 3$ nm and $R = 7$ nm), the square lattice (panels a and b) exhibits consistently higher transmission compared to the triangular lattice (panels c and d) as $V_0$ increases. This indicates that wave packets experience stronger scattering in the triangular configuration, likely due to the denser and more spatially compact arrangement of barriers, which increases the probability of interaction and reflection. The reduced transmission in the triangular case suggests that even without defects, lattice geometry plays a critical role in modulating the transport characteristics of Dirac fermions in graphene.

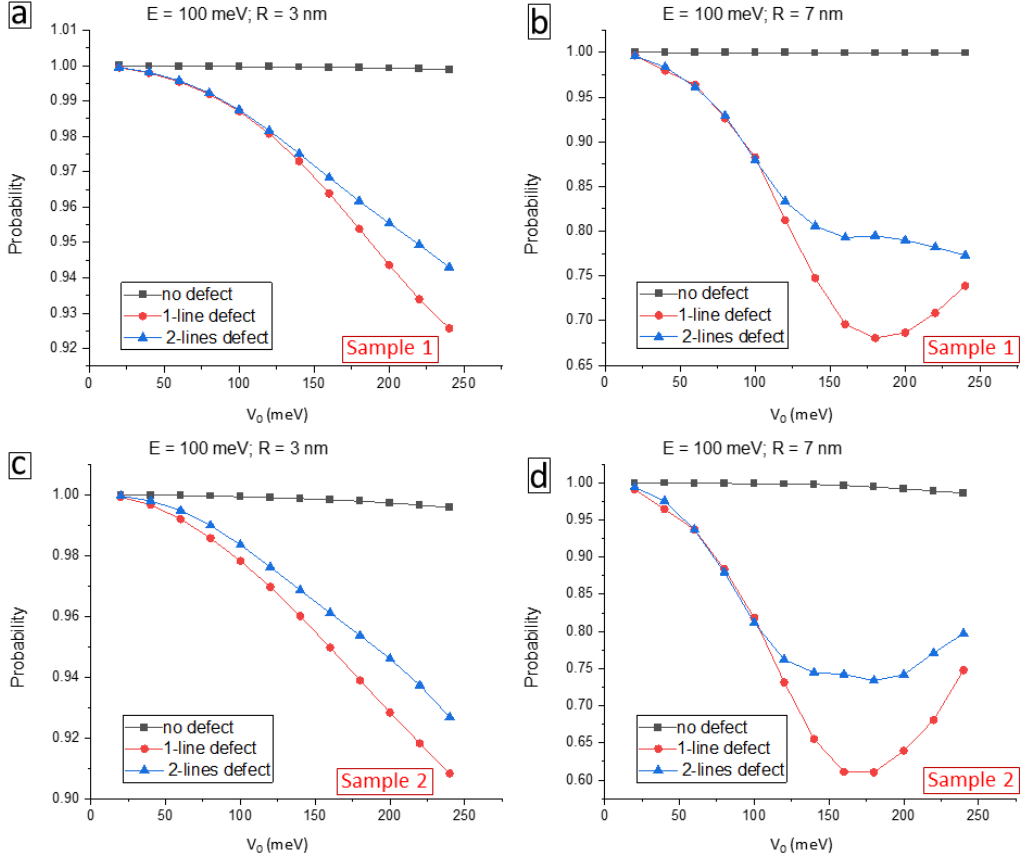

**Figure 2.** Transmission probability as a function of the barrier height $V_0$ for **Sample 1** (a, b) and **Sample 2** (c, d), for a wave packet with energy E=100 meV and spatial width $a_x$=10 nm. The radii of the circular potential barriers are R=3 nm (a, c), and R=7 nm (b, d). The black lines correspond to lattices without any defects, while the red and blue lines show results for lattices with one and two lines of removed barriers, respectively.

The results corresponding to the removal of a single line of potential barriers are shown by the red curves in Figure 2. Interestingly, the transmission probability decreases significantly compared to the defect-free case, despite the fact that removing scatterers would intuitively be expected to reduce scattering and enhance transmission. This counterintuitive behavior suggests that introducing a linear defect disturbs the periodic potential landscape in a way that enhances wave packet scattering, possibly due to interference effects or mode mismatches at the defect boundary. The impact of this effect also depends on the size of the potential barriers. For smaller barriers ($R = 3$ nm, panels a and c), the transmission decreases monotonically as the barrier height increases. In contrast, for larger barriers ($R = 7$ nm, panels b and d), the transmission exhibits non-monotonic behavior: it decreases initially, reaching a minimum, and then increases at higher potential values. This behavior may be attributed to partial transparency or tunneling resonances that emerge at specific barrier strengths.

Moreover, the suppression of transmission due to the single-line defect is more pronounced in the triangular lattice compared to the square lattice. This again points to the denser packing and more efficient scattering in the triangular geometry, which amplifies the influence of the defect on wave packet propagation.

When two lines of potential barriers are removed—results shown by the blue curves in Figure 2—the transmission probability increases compared to the single-line defect case. This trend is observed across both square and triangular lattices and for both barrier sizes ($R = 3$ nm and $R = 7$ nm), indicating that the removal of additional scattering regions eventually facilitates easier propagation of the wave packet. This behavior contrasts with the initial suppression observed for the single-line defect and highlights the nonlinear relationship between defect density and transmission: while partial disruption of the lattice increases scattering, further removal of barriers can create wider, less obstructed paths that enhance transport.

Additional features are worth noting. For small-radius barriers ($R = 3$ nm, panels a and c), the transmission with two-line defects decreases more slowly with increasing barrier height compared to the one-line defect case, indicating improved robustness of the transmission channel. In the case of larger barriers ($R = 7$ nm, panels b and d), the transmission also exhibits a non-monotonic trend, with a minimum followed by a noticeable recovery at higher barrier strengths. This suggests that for wider barriers, the system allows the formation of effective transmission corridors when defects are extended, possibly resembling tunneling channels or collimated wave paths.

Once again, the transmission remains lower for the triangular lattice than for the square lattice, reflecting the stronger intrinsic scattering associated with the denser arrangement. Nonetheless, the enhancement due to additional defect lines is clearly visible in both geometries, underlining the sensitive interplay between periodicity, defect structure, and wave dynamics in graphene.

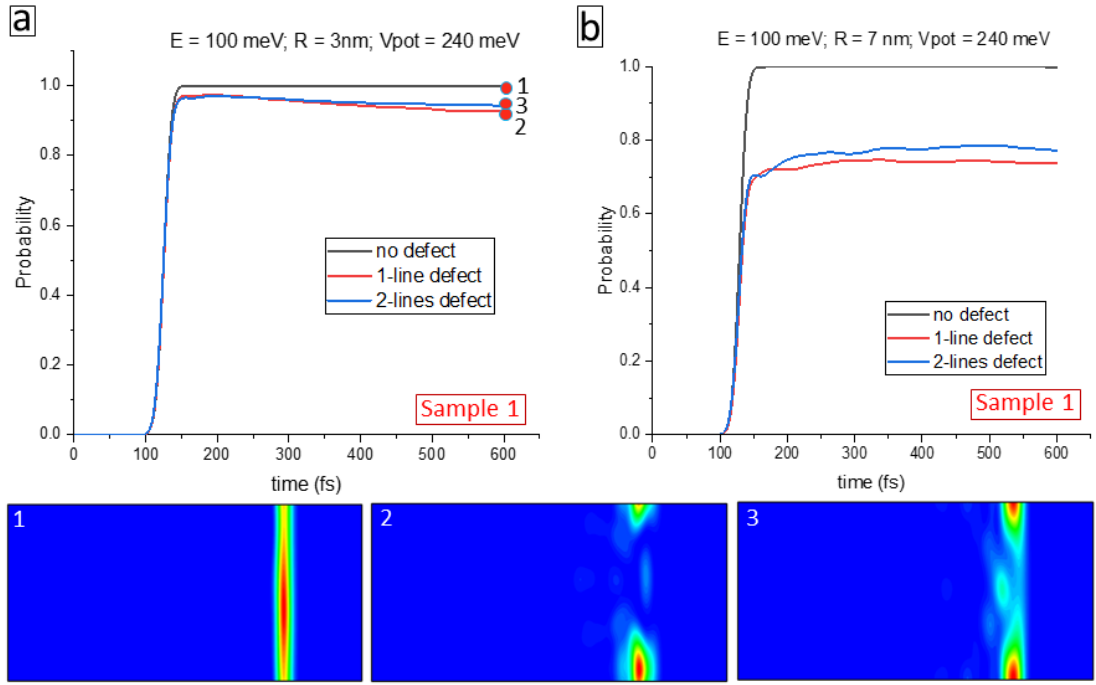

**Figure 3.** Time-dependent transmission probability for **Sample 1** with circular potential barriers of radius (a) $R = 3$ nm and (b) $R = 7$ nm. The incident wave packet has energy $E = 100$ meV and spatial width $a_x = 10$ nm. The potential height is fixed at $V_0 = 240$ meV. The black curve corresponds to a lattice without defects, while red and blue lines represent square or triangular arrangements with one- and two-line defects, respectively. Panels 1–3 display the corresponding wave packet snapshots at $t = 600$ fs for the cases labeled in (a).

To better understand the nonlinear dependence of wave packet transmission on the number of removed barrier lines, we analyze the time evolution of the transmission probability, as shown in Figure 3 for the square lattice configuration (**Sample 1**). The results are presented for two barrier sizes: $R = 3$ nm (panel a) and $R = 7$ nm (panel b), with a fixed potential height of $V_0 = 240$ meV and wave packet energy $E = 100$ meV. For the smaller barrier radius ($R = 3$ nm), the transmission rapidly rises and saturates close to unity in all three cases: no defect (black), one-line defect (red), and two-line defect (blue). This indicates minimal long-term scattering for smaller barriers, though a slight suppression is still observed when defects are present. The snapshot images at $t = 600$ fs (panels 1–3) provide additional insight into the spatial profile of the wave packet. In the defect-free case (panel 1), the wave packet remains highly collimated and undisturbed. However, the presence of defects (panels 2 and 3) causes noticeable distortion and lateral spreading, highlighting enhanced scattering even when fewer barriers are present. For the larger barrier radius ($R = 7$ nm), the differences are more pronounced. The black curve (no defect) shows an early and steady rise to a high transmission value, but the red and blue curves (one- and two-line defects) display a significantly slower increase and lower saturation levels. Interestingly, the two-line defect (blue) curve eventually exceeds the one-line defect (red), confirming the nonlinear behavior observed in Figure 2. This can be attributed to the formation of wider, more effective propagation channels as additional barriers are removed.

Overall, the time-resolved analysis supports the interpretation that structural defects can both hinder and enhance transmission depending on their extent, with barrier size playing a key role in shaping this behavior.

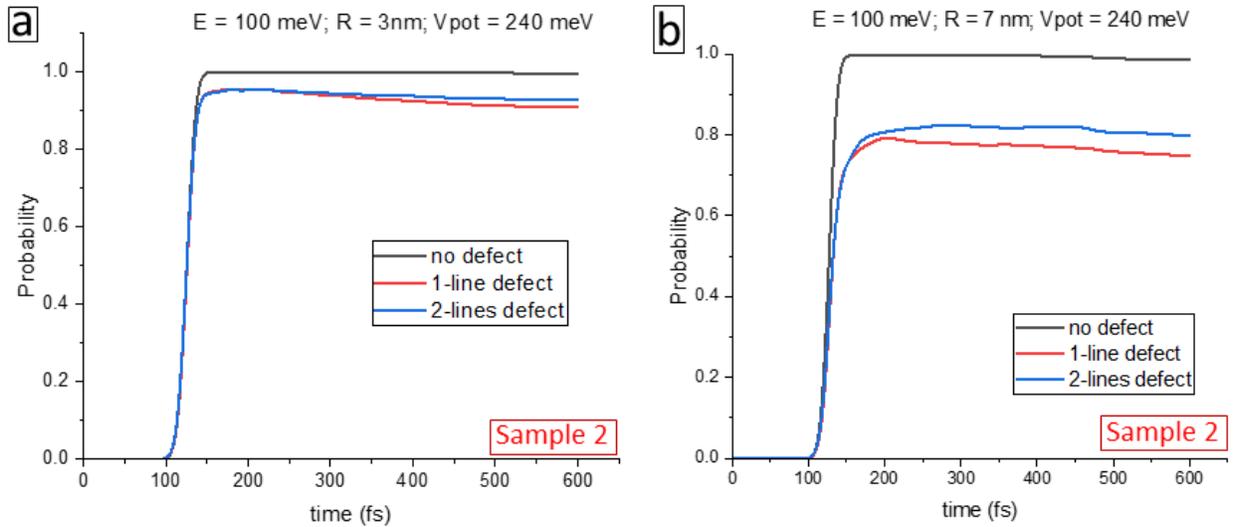

**Figure 4.** Transmission probability for **Sample 2** as a function of time for circular potential barriers with radius (a) $R = 3$ nm and (b) $R = 7$ nm. The wave packet energy is $E = 100$ meV, the potential height is $V_0 = 240$ meV, and the wave packet width is $a_x = 10$ nm. Black, red, and blue curves correspond to the cases of a defect-free lattice, and lattices with one-line and two-line defects, respectively.

A similar analysis was carried out for the triangular lattice arrangement, as shown in Figure 4, where the time-dependent transmission probabilities are plotted for **Sample 2**. As in the square lattice case, the results here are presented for barrier radii $R = 3$ nm (panel a) and $R = 7$ nm (panel b), with fixed potential height $V_o = 240$ meV and wave packet energy $E = 100$ meV.

The overall trends closely resemble those observed in the square lattice configuration (Figure 3), confirming that the nonlinear dependence of transmission on the number of barrier lines is a general feature of the system, not limited to lattice geometry. Specifically, for $R = 3$ nm (panel a), the transmission rises sharply and saturates near unity for the defect-free case (black curve), while the one-line (red) and two-line (blue) defect cases show slightly reduced and closely grouped transmission values. This suggests that for small barriers, the triangular arrangement still allows reasonably efficient transmission, though the tighter packing compared to the square lattice leads to more pronounced scattering, especially in the presence of defects. For the larger barrier radius ($R = 7$ nm, panel b), the differences become more distinct. As in the square lattice case, the one-line defect results in a notable suppression of transmission, while the two-line defect leads to a partial recovery, with the blue curve overtaking the red. The saturation values remain lower than those for the square lattice due to the increased density of scatterers in the triangular configuration, which enhances multiple scattering and wave packet dispersion.

Overall, the time-resolved analysis presented in Figures 3 and 4 confirms that the nonlinear influence of linear defects on wave packet transmission is a robust feature across both square and triangular lattice geometries. Structural defects can either hinder or enhance transmission depending on their extent: the removal of a single line of barriers leads to increased scattering and reduced transmission, while further removal creates wider propagation paths that partially restore transmission. The barrier size also plays a critical role in modulating this behavior, with larger scatterers amplifying the sensitivity of the system to defect-induced changes.

## CONCLUSIONS

We have employed the Dirac continuum model to study the propagation of wave packets in monolayer graphene with circular potential barriers arranged in square and triangular lattices, including cases with one- or two-line defects. We found that transmission exhibits a nonlinear dependence on the number of removed barrier lines: it initially decreases with the removal of a single line but increases again when more lines are removed. This effect is observed in both lattice types, with stronger scattering in the triangular geometry. Time-resolved analysis confirmed that barrier size and defect configuration significantly influence wave packet dynamics.

These results highlight the potential of structural design in controlling electron transport in graphene.